\documentclass[
    prb,
    aps,
    twocolumn,
    superscriptaddress,
    nofootinbib,
]{revtex4-1}

\usepackage{amsmath}
\usepackage{amssymb}
\usepackage{amsfonts}
\usepackage{bm}
\usepackage{bbm}
\usepackage{braket}
\usepackage{comment}
\usepackage{dcolumn}
\usepackage{enumerate}
\usepackage{epsfig}
\usepackage{gensymb}
\usepackage{graphicx}
\usepackage{indentfirst}
\usepackage{mathtools}
\usepackage{psfrag}
\usepackage{pst-all}
\usepackage{soul}
\usepackage{siunitx}
\usepackage{xcolor}
\usepackage{microtype}
\usepackage{txfonts}
\usepackage{txfontsb}
\usepackage{soul}
\usepackage{float} 

\usepackage[
colorlinks,
linkcolor=blue,
citecolor=blue,
urlcolor=blue]{hyperref} 
\usepackage[T1]{fontenc} 

                                      \def\pb{{\bf p}}               
                
\begin{document} 
\def\bibsection{\section*{\refname}}

\title{
Atomically thin silver films for enhanced nanoscale nonlinear optics
}

\author{Philipp~K.~Jenke}
\thanks{These authors contributed equally.}
\affiliation{University of Vienna, Faculty of Physics, Vienna Center for Quantum Science and Technology (VCQ), Boltzmanngasse 5, 1090 Vienna, Austria}
\affiliation{University of Vienna, Vienna Doctoral School in Physics, Boltzmanngasse 5, 1090 Vienna, Austria}

\author{Saad~Abdullah}
\thanks{These authors contributed equally.}
\affiliation{ICFO-Institut de Ciencies Fotoniques, The Barcelona Institute of Science and Technology, 08860 Castelldefels (Barcelona), Spain}

\author{Andrew~P.~Weber}
\affiliation{ICFO-Institut de Ciencies Fotoniques, The Barcelona Institute of Science and Technology, 08860 Castelldefels (Barcelona), Spain}

\author{\'{A}lvaro~Rodr\'{i}guez~Echarri}
\affiliation{Max-Born-Institut im Forschungsverbund Berlin e.V., Max-Born-Straße 2a, 12489 (Berlin), Germany}
\affiliation{Center for Nanophotonics, NWO Institute AMOLF, 1098 XG Amsterdam, The Netherlands}

\author{Fadil~Iyikanat}
\affiliation{ICFO-Institut de Ciencies Fotoniques, The Barcelona Institute of Science and Technology, 08860 Castelldefels (Barcelona), Spain}

\author{Vahagn~Mkhitaryan}
\affiliation{ICFO-Institut de Ciencies Fotoniques, The Barcelona Institute of Science and Technology, 08860 Castelldefels (Barcelona), Spain}

\author{Frederik~Schiller}
\affiliation{Centro de Física de Materials CSIC/UPV-EHU-Materials Physics Center, 20018 San Sebastián, Spain}
\affiliation{Donostia International Physics Center (DIPC), 20018 Donostia-San Sebastián, Spain}

\author{J. Enrique~Ortega}
\affiliation{Centro de Física de Materials CSIC/UPV-EHU-Materials Physics Center, 20018 San Sebastián, Spain}
\affiliation{Donostia International Physics Center (DIPC), 20018 Donostia-San Sebastián, Spain}
\affiliation{Departmento de Física Aplicada, Universidad del País Vasco, 20018 San Sebastián, Spain}

\author{Philip~Walther}
\affiliation{University of Vienna, Faculty of Physics, Vienna Center for Quantum Science and Technology (VCQ), Boltzmanngasse 5, 1090 Vienna, Austria}
\affiliation{University of Vienna, Research Platform for Testing the Quantum and Gravity Interface (TURIS), Boltzmanngasse 5, 1090 Vienna, Austria}
\affiliation{Christian Doppler Laboratory for Photonic Quantum Computer, Faculty of Physics, University of Vienna, 1090 Vienna, Austria}

\author{F.~Javier~Garc\'{i}a~de~Abajo}
\thanks{javier.garciadeabajo@nanophotonics.es}
\affiliation{ICFO-Institut de Ciencies Fotoniques, The Barcelona Institute of Science and Technology, 08860 Castelldefels (Barcelona), Spain}
\affiliation{ICREA-Institucio Catalana de Recerca i Estudis Avancats, Passeig Lluis Companys 23, 08010 Barcelona, Spain}

\author{Lee~A.~Rozema}
\thanks{lee.rozema@univie.ac.at}
\affiliation{University of Vienna, Faculty of Physics, Vienna Center for Quantum Science and Technology (VCQ), Boltzmanngasse 5, 1090 Vienna, Austria}

\date{\today}

\begin{abstract}{\bf \noindent
The inherently weak nonlinear optical response of bulk materials remains a fundamental limitation in advancing photonic technologies. Nanophotonics addresses this challenge by tailoring the size and morphology of nanostructures to manipulate the optical near field, thus modulating the nonlinear response. Here, we explore a complementary strategy based on engineering the electronic band structure in the mesoscopic regime to enhance optical nonlinearities. Specifically, we demonstrate an increase in second-harmonic generation (SHG) from crystalline silver films as their thickness is reduced down to just a few atomic monolayers. Operating at the boundary between bulk and two-dimensional systems, these ultra-thin films exhibit a pronounced enhancement of SHG with decreasing thickness. This enhancement stems from quantum confinement effects that modify the interaction between electronic states and incident light, which we explain based on quantum-mechanical calculation. Our atomically-thin crystalline silver films provide a new means to overcome the small interaction volumes inherent to nanophotonic platforms, enabling efficient nanoscale nonlinear optics with potential applications in photonics, sensing, and quantum technologies.
}\end{abstract}

\maketitle

\section*{Introduction}

Optical nonlinear processes are central to a wide range of applications in photonics~\cite{Davoyan2008}, medicine and biology~\cite{Fabrizio2016}, material science~\cite{Prylepa2018}, ultra-fast optics~\cite{Saltarelli2018}, and quantum technologies~\cite{Wagenknecht2010}. These applications typically rely on bulk nonlinearities in crystals~\cite{Franken1961, Dmitriev1999} or gases~\cite{Finn1971}, and require macroscopic interaction lengths to accumulate a substantial effect~\cite{Boyd}. In addition, care must be taken in tailoring the properties of the medium to accurately satisfy phase-matching conditions~\cite{Boyd}.
In contrast, nonlinear optics in ultra-thin films offers the potential for integration with compact photonic devices and allows for enhanced field confinement and stronger light–matter interactions over nanoscale volumes, reducing the reliance on phase matching and enabling new regimes of nonlinear light generation.

In recent years, following a nanophotonics approach, great progress has been made in reducing the volume of the nonlinear medium while reaching an efficient nonlinear response by controlling light at the subwavelength scale~\cite{Novotny2012, Husu2012, Zhang2024}. This has been accomplished through extrinsic manipulations such as plasmonic resonances~\cite{Kauranen2012, Mkhitaryan2023, Zhang2024},
strong electric field confinement~\cite{Bouhelier2003}, and engineered metasurfaces~\cite{Husu2012, Lee2014, Minovich2015}.
Aside from these extrinsic manipulations, electronic band structure engineering can be used to precisely tailor the intrinsic electronic and optical properties of a material. 
While such intrinsic manipulation is widely employed in platforms such as semiconductors~\cite{Lee2014}, semiconductor quantum dots~\cite{GarcadeArquer2021}, low-dimensional materials~\cite{AlonsoCalafell2020}, and ultra-cold atoms~\cite{Greiner2002}, it is novel in the field of nonlinear nanophotonics and plasmonics. 
Moreover, the small interaction volume regime provides a natural setting, with a largely untapped potential, to consider the application of established techniques, such as electronic or magnetic field manipulation, atomic doping, or stress tuning, to control and enhance the  nonlinear optical response.

Ultra-thin crystalline silver films are an exciting material platform where nanophotonic light control can be combined with mesoscopic band structure engineering~\cite{AbdElFattah2019, Echarri2021}. 
These films can support extremely confined plasmonic modes in the visible and near-infrared regimes~\cite{Mkhitaryan2023}, and intrinsic inelastic optical losses can be minimized to obtain spectrally narrow plasmons by maintaining a high-quality crystalline structure~\cite{McPeak2015, AbdElFattah2019,Mkhitaryan2023}.
Moreover, quantum-well states (QWS) are observed in the electronic band structure when the thickness approaches a few atomic layers~\cite{Kirilyuk1996, Pedersen1999, Hirayama2001, AbdElFattah2019, Echarri2021}, as experimentally confirmed by photoemission~\cite{Wachs1986} and electron tunneling~\cite{Jaklevic1971}. These QWS are sensitive to atomic-scale thickness variations~\cite{Pedersen1999, Hirayama2001} and can have a direct impact on the optical properties of atomically thin films. In particular, second-harmonic generation (SHG) has been predicted to oscillate as a function of the number of atomic layers~\cite{Kirilyuk1996, Pedersen1999, Hirayama2001}.

\begin{figure*}[t]
\centering
\includegraphics[width=0.85\textwidth]{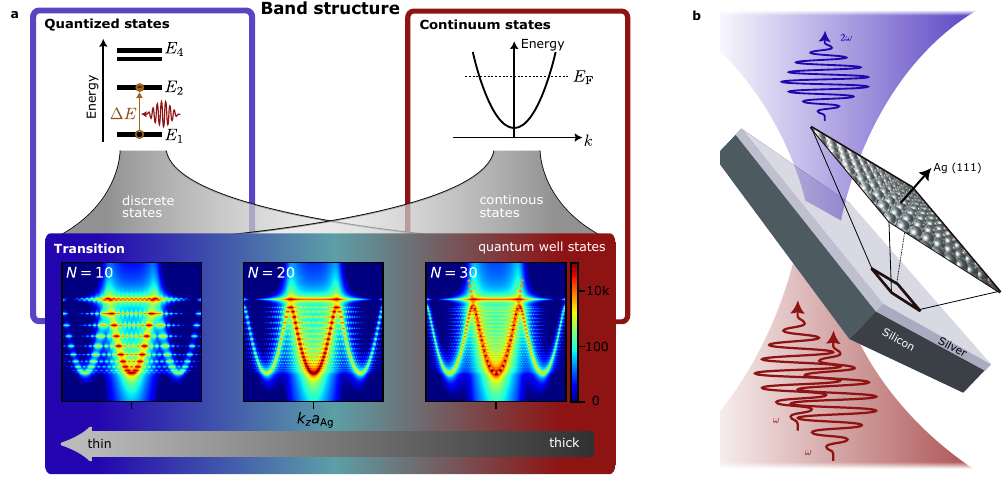}
\caption{\textbf{Interplay between electronic structure and nonlinear response in ultra-thin metal films.}
\textbf{a,}~The electronic density of states, represented here as a function of in-plane wave vector $k_z$ (bottom, horizontal scale) and energy (vertical state) shows an evolution from discrete quantum-well states (QWS) in thin films ($N=10$ atomic monolayers (ML)) to a coalescing band at larger thicknesses ($t=30\,\mathrm{ML}$). The discreteness of the electronic structure in the thin-film regime (upper-left scheme) enhances the nonlinear response (anharmonic excitation in response to harmonic light fields), in contrast to the more harmonic response associated with the parabolic band structure in the bulk limit (upper-right scheme). Here, $a_\mathrm{Ag}=2.36\,\mathrm{\AA}$ represents the atomic layer spacing.
\textbf{b,}~Schematic of the optical transmission geometry used to characterize the SHG signal. The ultra-thin silver film is excited from the silicon substrate, and SHG is collected in transmission. The silver film is a single-crystal with (111) orientation grown on the far silicon surface. A protecting silica capping layer covers the silver surface (not shown).
}
\label{fig:fig1}
\end{figure*}

In this work, we explore the strong thickness-dependent nonlinear response of atomically thin crystalline silver films to enhance the SHG efficiency by almost two orders of magnitude. In particular, for film thicknesses $t \lesssim 30\,$monolayers (MLs), quantum effects dominate and segregate electronic states into discrete quantized levels, akin to semiconductor quantum dots, as shown in Fig.~\ref{fig:fig1}a~\cite{Hirayama2001, Echarri2021}. Electrons are then confined to the QWS, resulting in discrete bands~\cite{Wachs1986} with less out-of-plane harmonic motion than in thicker crystalline structures, where electrons exhibit a parabolic out-of-plane dispersion relation.
The anharmonic motion associated with the confinement of QWS translates into stronger nonlinear response while their bulk metal counterparts require stronger field strengths to reach a similar level of nonlinear response~\cite{Echarri2021}.

Second-order nonlinearities at metallic interfaces were reported shortly after their observation in bulk crystals~\cite{Brown1965, Bloembergen1968, Franken1961}, and have since been extensively studied~\cite{Rudnick1971, Chen1981, Coutaz1987, Sipe1987, Wong1993, Chang1997, Pedersen1999, Hirayama2001}.
However, they generally exhibit much lower nonlinear conversion efficiency compared to nonlinear crystals such as lithium niobate or borate compounds. This is primarily due to the vanishing bulk second-order susceptibility in centrosymmetric metals like silver, which has a face-centered cubic structure~\cite{Boyd, Sipe1987}. In such materials, second-order responses predominantly arise from symmetry breaking at surfaces or interfaces.
However, when the interaction volume is constrained to be much less than the involved wavelengths (i.e., as it approaches the atomic scale), band structure engineering through the manipulation of surface properties, crystalline structure, and {the use of nanostructures} can dominate the linear and nonlinear optical responses~\cite{Kirilyuk1996, Echarri2021}. 
Modifying such parameters in the mesoscopic regime has been proposed to produce significant nonlinear efficiency enhancements~\cite{Echarri2021}, potentially strong enough to compensate for the reduction of interaction volume.

Here, we demonstrate that QWS in wafer-scale ultra-thin crystalline silver can enhance SHG by nearly two orders of magnitude while simultaneously reducing the interaction nonlinear volume. We have verified this observation across several excitation and SHG frequencies from the infrared to the mid-infrared range, quantifying the second-order nonlinear susceptibility. We have successfully modeled these effects based on a quantum mechanical description of the nonlinear response of silver film, going beyond previous classical nonlinear theories~\cite{Shen1989, Sipe1987}. By producing large-area crystalline films, we minimize inelastic losses, which are detrimental for SHG when decreasing the film thickness~\cite{Chang1997}. 
In our thin-crystalline-metal platform, light-matter interaction could be further increased by introducing nanostructures for local field enhancement~\cite{Simon1974,Bouhelier2003, Echarri2018}, offering a new route for highly efficient nonlinear nanophotonics and plasmonics~\cite{Echarri2021}.

\begin{figure}[t]
\centering
\includegraphics[width=0.45\textwidth]{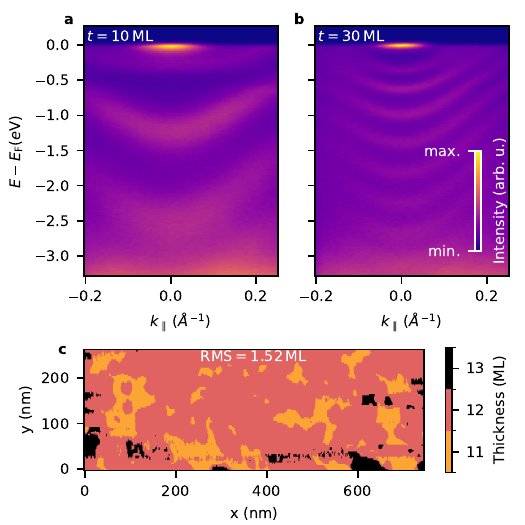}
\caption{\textbf{Film characterization.}
\textbf{a/b,}~Angle-resolved photoemission spectroscopy images of crystalline silver films with (111)~crystal orientation showing QWS. The photoemission intensity is plotted relative to the Fermi energy $E_\mathrm{F}$ as a function of in-plane electron wave vector $k_\parallel$. The film thickness $t$ is expressed in terms of the number of (111) atomic monolayers (MLs): \textbf{a,}~$10\,\mathrm{ML}$ and \textbf{b,}~$30\,\mathrm{ML}$. \textbf{c,}~Scanning tunneling microscope (STM) image over an extended area of a crystalline silver film with a root mean square roughness (RMS) of $1.52\,\mathrm{ML}$.
}
\label{fig:fig2}
\end{figure}

\begin{figure*}[t]
\centering
\centering
\includegraphics[width=0.85\textwidth]{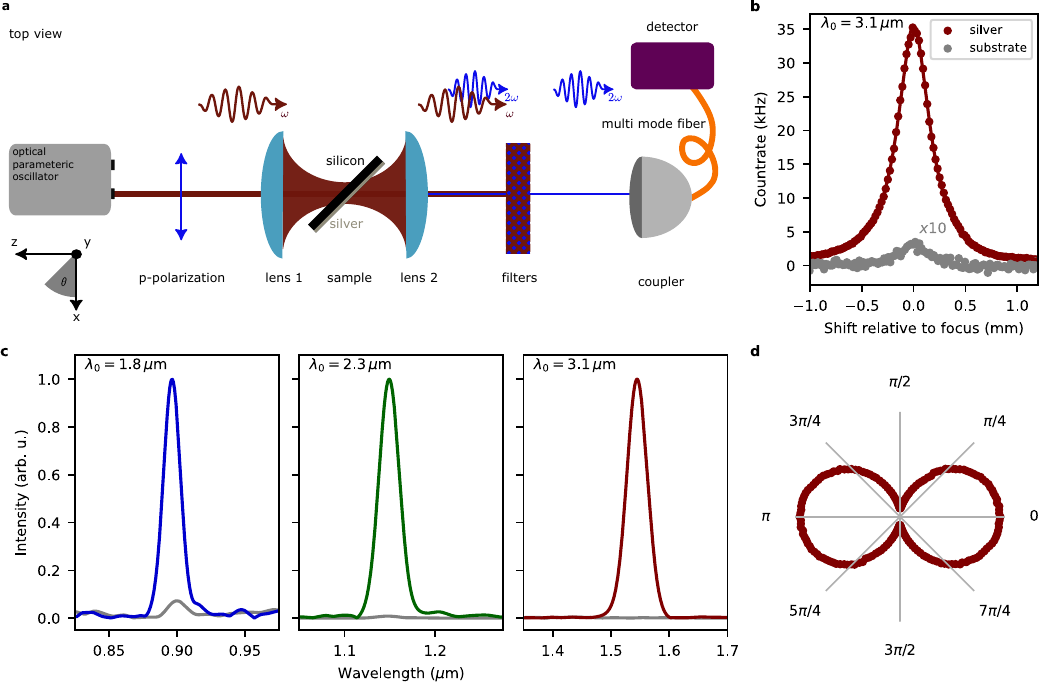}
\caption{\textbf{Characterization of the SHG signal.}
    \textbf{a,}~Sketch of the optical transmission setup: A Ti:Sapphire oscillator (not shown) pumps an optical parametric oscillator providing p-polarized long excitation pulses ($200\,\mathrm{fs}$), which are focused via the first lens 1 into the $45\,^\circ$ tilted sample. The generated SHG signal is collimated by lens 2, and optical filters spectrally separate it from the excitation light. The isolated SHG signal is then coupled into a multi-mode fiber and directed to the detector.
    \textbf{b,}~Detected SHG count rate from two different lateral sample positions corresponding to the bare silicon substrate (multiplied by $10$) and the deposited silver film in gray and red, respectively.
    \textbf{c,}~Spectra of the SHG signal for different excitation wavelengths.
    %
    %
    {\textbf{d,}~SHG for fixed p-polarization excitation and different angles of the polarizer to the right of the sample, with $0\,^\circ$ corresponding to p-polarization. The silver film is $17\,$ML thick in \textbf{b--d}.}
}
\label{fig:fig3}
\end{figure*}

\begin{figure*}[t]
\centering
\includegraphics[width=0.8\textwidth]{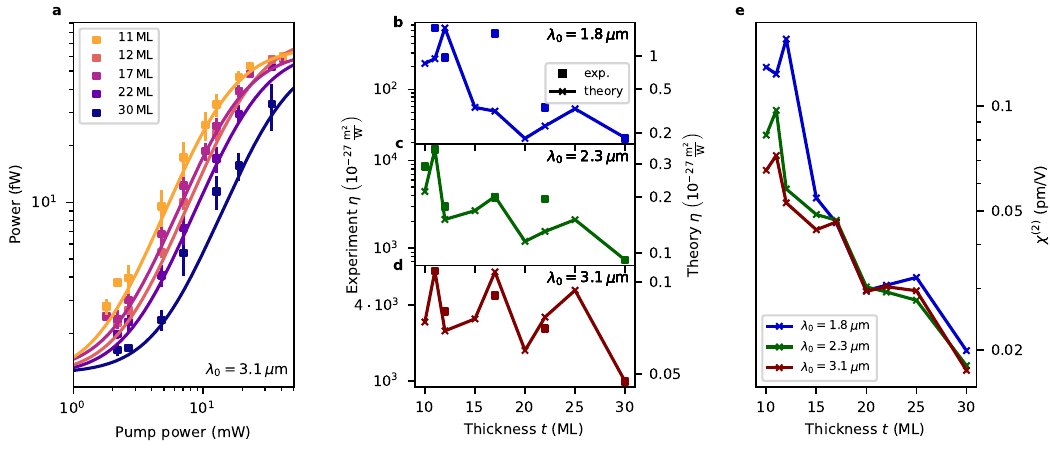}
\caption{
    \textbf{SHG efficiencies for different silver film thicknesses.}
    \textbf{a,}~Power scaling of the measured SHG signal (symbols) in crystalline silver films of different thicknesses for a fixed incident wavelength of $3.1\,\mu\mathrm{m}$. A second-order power law is used to fit the data, including a constant background and detector saturation.
    \textbf{b-d,}~Absolute SHG conversion efficiencies $\eta$ for various excitation wavelengths $\lambda_0$ and film thicknesses. Experimental efficiencies (squares) refer to peak intensities and are extracted from the power scaling in \textbf{a} without normalization. Theoretical values are estimated from the SHG susceptibility in \textbf{e}.
    \textbf{e,}~Silver film SHG susceptibility for an intrinsic damping $\gamma=0.1\,\mathrm{eV}$.
}
\label{fig:fig4}
\end{figure*}

\section*{Results}

\subsection*{Ultra-thin-film platform}

Our material platform consists of crystalline silver films with $(111)$ surface orientation and $10$ to $30$ atomic layers in thickness. The individual crystal domains may extend over a few square centimeters~\cite{AbdElFattah2019}. This large-area crystallinity is crucial for the formation of high-quality QWS, providing very low sheet resistance~\cite{AbdElFattah2019}, low inelastic losses, and high-quality plasmonic excitations, while remaining compatible with nanolithography and other nanophotonic manipulation techniques~\cite{Rodionov2019, Mkhitaryan2023}. We employ epitaxial crystalline growth of silver on a crystalline silicon wafer (see Methods). This process is scalable and without lateral restrictions while preserving atomic-scale precision and low surface roughness (Fig.~\ref{fig:fig2}c).
Our approach is more flexible than previously demonstrated wet-chemistry synthesis methods, in which the lateral film size is correlated to the thickness~\cite{Pan2024}.

As discussed in the Methods and Supplementary Information (SI), our silver films are grown in ultra-high vacuum (UHV), and characterized in-situ~\cite{AbdElFattah2019}. 
We first use angle-resolved photoemission spectroscopy (ARPES) to verify the film thickness by detecting and characterizing QWS. The observed changes in the band structure as a function of the film thickness are illustrated in Fig.~\ref{fig:fig2}a/b. We then use scanning tunneling microscopy (Fig.~\ref{fig:fig2}c) to confirm the crystalline surface topography, indicating that we can produce flat films with atomic-scale terrace steps.
Finally, before removal from UHV, our silver films are coated with a thin ($\approx1\,\mathrm{nm}$) silicon passivation layer~\cite{AbdElFattah2019, Mkhitaryan2023} to ensure their stability under ambient conditions.
We then remove the films from UHV and measure the SHG response for different film thicknesses (Fig.~\ref{fig:fig1}b) to probe the optical nonlinear response.

\subsection*{Measured SHG signal}

To isolate the nonlinear SHG signal from the silver films, we use a modified focus scan setup (Fig.~\ref{fig:fig3}a). In our configuration, the $45\,^\circ$-tilted sample is scanned through the focal point of a femtosecond-pulsed mid-infrared pump beam (with a center wavelength $\lambda_0$ of $1.8$, $2.3$ or $3.1\,\mu\mathrm{m}$). The resulting SHG signal (at $0.9$, $1.15$ or $1.55\,\mu\mathrm{m}$, respectively) is measured in transmission as detailed in Methods. To avoid absorption of the SHG signal by the silicon substrate, the sample is oriented so that the pump traverses the substrate before interacting with the silver film. All optical measurements are conducted under ambient pressure and temperature conditions within several days after fabrication, with no sample degradation observed for weeks (SI Fig.~S6
). This underscores the efficacy of the passivation layer, as well as the durability and resilience of the fabricated films.

A representative focus scan measurement is presented in Fig.~\ref{fig:fig3}b, demonstrating notable SHG signal strengths only at the focal point with the highest fluence (red data), and essentially no signal from the substrate (grey points).
In Fig.~\ref{fig:fig3}c, we plot the spectra of the measured nonlinear signal at the focal point for pump wavelengths of $1.8$, $2.3$, and $3.1\,\mu\mathrm{m}$. 
These spectra reveal clear peaks at the expected SHG wavelengths, verifying the nonlinear process spectrally.

We next study the polarization behavior of the nonlinear signal.
{The generated SHG is co-polarized with the incoming p-polarized light, as shown in Fig.~\ref{fig:fig3}d. This strongly polarized signal indicates a coherent nonlinear process. We did not perform a study of the SHG dependence on the linear polarization of the excitation, since excitation through the silicon substrate preserves only p-polarization.}

To quantify the conversion efficiency and confirm the second-order nature of the SHG signal, we measure the average SHG power $P_{\mathrm{SHG}}$ as a function of average excitation power $P_0$ at pulsed excitation.
Results for an excitation wavelength of $3.1\, \mu\mathrm{m}$ are presented in Fig.~\ref{fig:fig4}a for various silver film thicknesses. Analogous measurements for excitation wavelengths of $1.8$ and $2.3\,\mu\mathrm{m}$ are provided in SI Fig.~S5
. At low excitation powers, the detected signal is primarily limited by dark counts, while at high powers, detector saturation becomes significant.
In the intermediate regime, a quadratic dependence with a slope of $2$ on a log–log scale is observed, confirming the second-order power scaling characteristic of the SHG process.
As detailed in the Methods section, the regions affected by detector nonlinearity can be well-modeled, enabling the extraction of the actual average SHG power $P_{\mathrm{SHG}}$ from the measured average detector power $P_{\mathrm{det}}$.
The SHG conversion efficiency $\eta = I_{\mathrm{SHG,peak}} / I_{\mathrm{0,peak}}^2$, is the single free parameter used to fit each power dependence dataset in Fig.~\ref{fig:fig4}a. Here, $I_{\mathrm{peak}}$ denotes the peak intensity within the silver film, corrected for transmission losses in the setup. The results reveal that thinner silver films produce stronger absolute SHG signals than thicker ones.
Applying the analogue analysis used in Fig.~\ref{fig:fig4}a to other excitation wavelengths, we obtain the SHG conversion efficiencies shown in Fig.~\ref{fig:fig4}b--d. For all considered excitation wavelengths, we observe a clear increase in absolute SHG conversion efficiency with decreasing film thickness, even without any normalization.
From a standard nonlinear optics perspective, this trend is counterintuitive, as a quadratic increase in SHG intensity with interaction length would be expected for a non-phase-matched process with short propagation lengths~\cite{Boyd}.
In contrast, we observe the opposite trend: an increase in nonlinear conversion with decreasing thickness, caused by a more anharmonic electron motion associated with QWS (i.e., far from harmonic motion in a parabolic potential), as they become increasingly more separated in energy.
While previous work has reported oscillations in SHG with film thicknesses ~\cite{Pedersen1999, Hirayama2001}, here we instead fabricate and measure an integer number of layers, achieving an enhanced conversion efficiency of almost two orders of magnitude by decreasing the film thickness by a factor of~$3$. This provides a drastic boost in conversion efficiency for the first time in this system, showing the potential of band structure engineering in this novel nanophotonic platform.

\subsection*{Theoretical model}

To validate our experimental results, we model the silver layers as a continuous distribution of out-of-plane second-harmonic induced dipoles. We based our calculations on a quantum-mechanical description of QWS~\cite{Echarri2021} using a film-confining model potential, fitted to reproduce salient features of the silver surface electronic structure; a nonlinear extension of the random-phase approximation is then followed; although the model potential is taken to be translationally invariant along surface directions, we incorporating band-dependent effective electron masses from density-functional theory simulations including in-plane atomic corrugation (see Ref.~\cite{Echarri2021} and Methods). 

We compare the experimental and theoretical SHG conversion efficiencies $\eta$ in Fig.~\ref{fig:fig4}b-d for the considered excitation wavelengths of $1.8$, $2.3$, and $3.1\,\mu$m. The calculated second-order susceptibilities exhibit a marked increase as the number of layers decreases (Fig.~\ref{fig:fig4}e), in agreement with the trend observed experimentally. This indicates that the rise in second-order susceptibility is due to thickness-dependent changes in the electronic band structure of the material. However, the variation in calculated SHG conversion efficiencies with incident light wavelength differs from the experimental results. This discrepancy may be attributed to thickness-dependent changes in the nonlinear optical response due to ultra-fast thermal effects~\cite{RodrguezEcharri2023} or broadening effects, which were not included in our calculations.

\section*{Discussion}

We have demonstrated that crystalline ultra-thin silver films are a promising platform for nonlinear nanophotonics, holding the potential to enhance optical nonlinear interactions while simultaneously reducing the interacting volume. Due to the quantum-well band structure, ultra-thin films overcome the classical scaling of the nonlinear intensity with the interaction length, resulting in stronger nonlinear responses as the film thickness decreases. This approach to enhancing the nonlinear response is compatible with nanophotonic strategies such as embedding in optical cavities, which could work well in the visible and mid-infrared spectral ranges.

While the nonlinear conversion efficiencies observed in silver films are still lower than those in other 2D materials such as transition metal dichalcogenides, they offer advantages in terms of stronger optical response, more flexible nanopatterning, and higher thermal and electrical conductivities associated with their metallic nature. In particular, lateral patterning can enhance the linear response~\cite{Mkhitaryan2023}, and in turn, also the nonlinear behavior in virtue of the associated optical field confinement and enhancement. In this context, graphene plasmons have created great expectations because of their appealing properties~\cite{GarcadeAbajo2014}, including strong nonlinearities~\cite{Cox2019},
but they are constrained to the mid-infrared range. In contrast, silver films cover a plasmonic behavior extending to the near-infrared and visible domains~\cite{Echarri2021,Mkhitaryan2023}. Plasmon resonances, nanostructures, and electric field enhancements have the potential to boost the conversion efficiency to strengths comparable to those of 2D materials while maintaining electronic tunability~\cite{Zhang2024, Nahata2003, Metzger2015}. 
In addition, theory predicts even greater enhancements for film thicknesses below $10\,\mathrm{ML}$~\cite{Echarri2021}, offering exciting prospects as advances in fabrication continue to improve the stabilization of such ultra-thin films over large areas.

\section*{Methods}

\noindent\textbf{Experimental details.}
We conducted SHG measurements using a modified focus scan setup, where the SHG signal was measured as the sample was traversed along the optical axis through the focus of the excitation beam (Fig.~\ref{fig:fig3}a), while different lateral spots were excited by moving the tilted sample within the focus plane. Our incident light beam comprised linearly p-polarized gaussian pulses, typically $\tau=200\,\mathrm{fs}$ in duration, with a tunable center wavelength $\lambda_0$ ranging from $1.8\,\mu\mathrm{m}\ (0.69\,\mathrm{eV})$ to $3.1\,\mu\mathrm{m}\ (0.40\,\mathrm{eV})$ (Fig.~S7
, SI) at a repetition rate $f=76\,\mathrm{MHz}$. This beam was generated by an optical parametric oscillator optically pumped by a Ti:Sapphire oscillator. The pulse duration and repetition rate varied slightly due to daily alignment.

The maximum acquisition time for SHG measurements before readjusting the excitation power was around $20$ to $30\,\mathrm{min}$; during this time, we observed power fluctuations of $<1\,\%$. A half-wave plate was employed to rotate the polarization of the incoming beam, and linear polarization was ensured by a polarizer (p-polarization was used, unless otherwise stated). An optical system comprising two lenses, each with a $20\,\mathrm{mm}$ focal length, was used to focus the excitation beam onto the sample and collimate the signal of interest. The beam waist $\sigma_0$ for the excitation wavelengths of $1.8$, $2.3$, and $3.1\,\mu\mathrm{m}$ were $11.8$, $13.8$, and $17.0\,\mu\mathrm{m}$, respectively (Fig.~S9
, SI). The sample was excited from the back side to prevent the absorption of the nonlinear signal at the substrate. Most of the nonlinear emission was generated at the focus plane, where fluence was maximum, refracting the incident angle from $\theta_0=45\,^\circ$ in free space to $\theta_1\approx12\,^\circ$ inside the silicon substrate. After the collimation lens, the nonlinear signal was isolated by optical transmission filters and coupled into a multimode fiber before being detected with InGaAs or Si single-photon detectors. The latter was only used for $1.8\,\mu\mathrm{m}$ excitation wavelength. Excitation and detection efficiencies have been taken into account. We confirmed the spectral characteristics of the signal through spectral verification using an interferometer-based spectrometer from Nireos by Fourier transforming the autocorrelation function of the interferogram (Fig.~S10
, SI).

\noindent\textbf{Extracting the second-order nonlinear susceptibility.}
The SHG power, $P_\mathrm{SHG}$, was related to the excitation power $P_0$ by assuming a second-order power law:
\begin{align}
    P_\mathrm{SHG} = \eta P_0(\omega)^2, \label{eq:power_law}
\end{align}
where $\eta$ is the fitted conversion efficiency. In our analysis, we also account for a constant background offset $P_\mathrm{dark}$ and detector saturation effects, using the expression:
\begin{align}
    P_\mathrm{det} = \frac{P_\mathrm{SHG}}{1 + P_\mathrm{SHG}/P_\mathrm{sat}} + P_\mathrm{dark}, \nonumber
\end{align}
which allows us to extract $P_\mathrm{SHG}$ from the measured detector signal $P_\mathrm{det}$ and the fitted value of $P_\mathrm{sat}$.

The SHG conversion efficiency is definded by $\eta = I_{\mathrm{SHG,peak}} / I_{\mathrm{0,peak}}^2 $ and the peak intensity is given by
$
I_{\mathrm{peak}} = \frac{ 2 f_{\mathrm{shape}}}{f \tau A} P_{\mathrm{ave}},
$
where $A=\pi\sigma_0^2\sec\theta_1$ is the focus area and $f_{\mathrm{shape}}=94\%$ accounting for the gaussian pulse shape.

\noindent\textbf{Fabrication and characterization details.}
Si-passivated Ag films were prepared under ultra-high vacuum conditions on $525\,\mu\mathrm{m}$ thick Si(111) substrates following the procedures in Ref.~\cite{AbdElFattah2019}. The crystallinity, atomic flatness, and thickness of the Ag films were characterized by low-energy electron diffraction (Fig.~S1
, SI), ARPES (Figs.~\ref{fig:fig2}a-b and S2
, SI), STM (Fig.~\ref{fig:fig2}c), and X-ray photoelectron spectroscopy (XPS) (Fig.~S3
, SI) before removal from UHV. SEM (see Fig.~S4
, SI) was used to confirm the flatness and absence of dewetting in the passivated film after air exposure. Photoemission measurements were performed using a SPECS Phoibos $150$ electron analyzer. For ARPES measurements, a monochromatized He gas discharge lamp operating at the He I$\alpha$ excitation energy ($21.22\,\mathrm{eV}$) was used. A monochromatized Al K$\alpha$ source ($1486.71\,\mathrm{eV}$) was used for XPS measurements. The experimental energy resolution was better than $40\,\mathrm{meV}$ and $500$ meV for ARPES and XPS, respectively. The angular resolution for ARPES was better than $0.4^\circ$. STM data were collected using an Omicron VT setup. The sample was kept at a temperature of $100\,\mathrm{K}$ for the photoemission measurements and at room temperature ($300\,\mathrm{K}$) for the LEED, STM, and SEM measurements.  

\noindent\textbf{Density functional theory calculations.}
We conduct first-principles density functional theory (DFT) calculations on Ag(111) films of varying thicknesses. The Kohn-Sham one-electron electronic structure is computed using the Perdew-Burke-Ernzerhof (PBE) exchange-correlation functional~\cite{PBE96}, as implemented in the Vienna {\it ab initio} simulation package (VASP)~\cite{Kresse1993,Kresse1996}. The plane wave cutoff energy is set to $600\,\mathrm{eV}$. The experimentally determined bulk lattice constant ($4.09\,\mathrm{\AA}$) is used to construct thin silver films by fixing the interatomic bond distances. A $18\times18\times1$ $\Gamma$-centered $k$-point mesh is used for Brillouin zone sampling. Each Ag film is positioned at the center of the simulation cell, with a vacuum gap of $10\,\mathrm{\AA}$ in the vertical direction to avoid spurious interactions between adjacent cells. Based on the self-consistent charge density, we extract single-particle energies throughout the Brillouin zone to obtain the electronic band structures of the layers. Effective masses near the Fermi level are then determined using the methodology described in our previous work~\cite{Echarri2021}.

\noindent\textbf{Quantum-mechanical description of silver films.}
Following the prescription described in Ref.~\cite{Echarri2021}, we simulate the nonlinear optical response of few-atom-thick films of Ag(111) based on a single-particle wave function formalism, where electrons are subject to an atomic potential that reproduces the main features of the measured silver surface (work function, projectec gap, etc.). From the resulting thickness-dependent wave functions, we compute the induced dipole per unit area $\pb$ under excitation by an out-of-plane optical electric potential $\phi=-zE_0$ determined by the normal electric field amplitude $E_0$ under the illumination conditions. We solve the problem in one dimension and obtain a second-harmonic normal dipole per unit area, from which we determine the SHG susceptibility as $\chi^{(2)} = p_z/(t E_0^2)$, where $t$ is the thickness.

\section*{data availability}
The data and computer codes that support the plots within this paper and other findings of this study are available from the corresponding author upon reasonable request.

\section*{}
\bibliographystyle{naturemag}

\bibliography{references}

\section*{acknowledgments}
{
This project has received funding from the European Union (HORIZON Europe Research and Innovation Programme, EPIQUE, No 101135288), and the European Research Council (Advanced Grant 101141220-QUEFES).
Views and opinions expressed are however those of the author(s) only and do not necessarily reflect those of the European Union or the European Commission-EU. Neither the European Union nor the granting authority can be held responsible for them.
This research was also funded in whole or in part by the Austrian Science Fund (FWF)[10.55776/COE1] (Quantum Science Austria), [10.55776/F71] (BeyondC) and [10.55776/FG5] (Research Group 5).
For open access purposes, the author has applied a CC BY public copyright license to any author accepted manuscript version arising from this submission.
Funding was also received from 
the Spanish MICINN (PID2024-157421NB-I00 and Severo Ochoa CEX2019-000910-S), and the Catalan AGAUR (Grant No. 2024 FI-2 00052) and CERCA Programs.
P.W. acknowledges financial support from the Air Force Office of Scientific Research under award number FA9550-21-1-0355 (Q-Trust) and FA8655-23-1-7063 (TIQI), as well as the National Foundation for Research, Technology and Development and the Christian Doppler Research Association.
P.K.J. acknowledges support from the University of Vienna via the Vienna Doctoral School.
}



\end{document}